\documentclass[manuscript]{aastex}

\newcommand{\myemail}{ttsuka@mx.ibaraki.ac.jp}

\slugcomment{Draft version 1}

\shorttitle{High-Resolution Submillimeter and Near-Infrared Studies of the Transition Disk around \object{Sz 91}}
\shortauthors{T. Tsukagoshi et al.}

\begin{document}

\title{High-Resolution Submillimeter and Near-Infrared Studies of the Transition Disk around \object{Sz 91}}

\author{%
Takashi Tsukagoshi\altaffilmark{1},
Munetake Momose\altaffilmark{1},
Jun Hashimoto\altaffilmark{5},
Tomoyuki Kudo\altaffilmark{2},
Sean Andrews\altaffilmark{4},
Masao Saito\altaffilmark{2}, 
Yoshimi Kitamura\altaffilmark{3},
Nagayoshi Ohashi\altaffilmark{2},
David Wilner\altaffilmark{4},
Ryohei Kawabe\altaffilmark{2},
Lyu Abe\altaffilmark{7},
Eiji Akiyama\altaffilmark{2},
Wolfgang Brandner\altaffilmark{8},
Timothy D. Brandt\altaffilmark{9},
Joseph Carson\altaffilmark{10},
Thayne Currie\altaffilmark{28},
Sebastian E. Egner\altaffilmark{11},
Miwa Goto\altaffilmark{12},
Carol Grady\altaffilmark{13},
Olivier Guyon\altaffilmark{11},
Yutaka Hayano\altaffilmark{11},
Masahiko Hayashi\altaffilmark{2},
Saeko Hayashi\altaffilmark{11},
Thomas Henning\altaffilmark{8},
Klaus W. Hodapp\altaffilmark{14},
Miki Ishii\altaffilmark{2},
Masanori Iye\altaffilmark{2},
Markus Janson\altaffilmark{15},
Ryo Kandori\altaffilmark{2},
Gillian R. Knapp\altaffilmark{15},
Nobuhiko Kusakabe\altaffilmark{2},
Masayuki Kuzuhara\altaffilmark{2,16},
Jungmi Kwon\altaffilmark{27},
Mike McElwain\altaffilmark{17},
Taro Matsuo\altaffilmark{18},
Satoshi Mayama\altaffilmark{27},
Shoken Miyama\altaffilmark{19},
Jun-ichi Morino\altaffilmark{2},
Amaya Moro-Mart{\'i}n\altaffilmark{20},
Tetsuro Nishimura\altaffilmark{11},
Tae-Soo Pyo\altaffilmark{11},
Eugene Serabyn\altaffilmark{21},
Takuya Suenaga\altaffilmark{27},
Hiroshi Suto\altaffilmark{2},
Ryuji Suzuki\altaffilmark{2},
Yasuhiro Takahashi\altaffilmark{6},
Hideki Takami\altaffilmark{2},
Michihiro Takami\altaffilmark{22},
Naruhisa Takato\altaffilmark{11},
Hiroshi Terada\altaffilmark{11},
Christian Thalmann\altaffilmark{23},
Daigo Tomono\altaffilmark{11},
Edwin L. Turner\altaffilmark{9},
Tomonori Usuda\altaffilmark{11},
Makoto Watanabe\altaffilmark{24},
John P. Wisniewski\altaffilmark{25},
Toru Yamada\altaffilmark{26},
and Motohide Tamura\altaffilmark{2,6}
}

\altaffiltext{1}{College of Science, Ibaraki University, Bunkyo 2-1-1, Mito 310-8512, Japan: \myemail}
\altaffiltext{2}{National Astronomical Observatory Japan(NAOJ), Osawa 2-21-1, Mitaka, Tokyo 181-8588, Japan}
\altaffiltext{3}{Institute of Space and Astronautical Science, Japan Aerospace Exploration Agency, Yoshinodai 3-1-1, Sagamihara, Kanagawa 229-8510, Japan}
\altaffiltext{4}{Harvard-Smithsonian Center for Astrophysics, 60 Garden Street, Cambridge, MA 02138, USA}
\altaffiltext{5}{Department of Physics and Astronomy, The University of Oklahoma, 440 W. Brooks St. Norman, OK 73019, USA}
\altaffiltext{6}{School of Science, The University of Tokyo, Hongo 7-3-1, Bunkyo, Tokyo 113-0033, Japan}

\altaffiltext{7}{Lboratoire Lagrange (UMR 7293), Universit\'e de Nice-Sophia Antipolis, CNRS, Observatoire de la C\^ote d'Azur, 28 avenue Valrose, 06108 Nice Cedex 2, France}
\altaffiltext{8}{Max Planck Institute for Astronomy, K\"onigstuhl 17, 69117 Heidelberg, Germany}
\altaffiltext{9}{Department of Astrophysical Sciences, Princeton University, Peyton Hall, Ivy Lane, Princeton, NJ 08544, USA}
\altaffiltext{10}{Department of Physics and Astronomy, College of Charleston, 58 Coming St., Charleston, SC 29424, USA}
\altaffiltext{11}{Subaru Telescope, 650 North A'ohoku Place, Hilo, HI 96720, USA}
\altaffiltext{12}{Universit\"ats-Sternwarte M\"unchen, Ludwig-Maximilians-Universit\"at, Scheinerstr. 1, 81679 M\"unchen, Germany}
\altaffiltext{13}{Exoplanets and Stellar Astrophysics Laboratory, Code 667, Goddard Space Flight Center, Greenbelt, MD 20771 USA}
\altaffiltext{14}{Institute for Astronomy, University of Hawaii, 640 N. A'ohoku Place, Hilo, HI 96720, USA}
\altaffiltext{15}{Department of Astrophysical Sciences, Princeton University, Peyton Hall, Ivy Lane, Princeton, NJ 08544, USA}
\altaffiltext{16}{Department of Earth and Planetary Science, The University of Tokyo, 7-3-1 Hongo, Bunkyo-ku, Tokyo, 113-0033, Japan}
\altaffiltext{17}{Exoplanets and Stellar Astrophysics Laboratory, Code 667, Goddard Space Flight Center, Greenbelt, MD 20771 USA}
\altaffiltext{18}{Department of Astronomy, Kyoto University, Kitashirakawa-Oiwake-cho, Sakyo-ku, Kyoto, Kyoto 606-8502, Japan}
\altaffiltext{19}{Hiroshima University, 1-3-2, Kagamiyama, Higashihiroshima, Hiroshima 739-8511, Japan}
\altaffiltext{20}{Department of Astrophysics, CAB-CSIC/INTA, 28850 Torrej\'on de Ardoz, Madrid, Spain}
\altaffiltext{21}{Jet Propulsion Laboratory, California Institute of Technology, Pasadena, CA, 171-113, USA}
\altaffiltext{22}{Institute of Astronomy and Astrophysics, Academia Sinica, P.O. Box 23-141, Taipei 10617, Taiwan}
\altaffiltext{23}{Astronomical Institute "Anton Pannekoek", University of Amsterdam, Postbus 94249, 1090 GE, Amsterdam, The Netherlands}
\altaffiltext{24}{Department of Cosmosciences, Hokkaido University, Kita-ku, Sapporo, Hokkaido 060-0810, Japan}
\altaffiltext{25}{Department of Astronomy, University of Washington, Box 351580 Seattle, WA 98195, USA}
\altaffiltext{26}{Astronomical Institute, Tohoku University, Aoba-ku, Sendai, Miyagi 980-8578, Japan}
\altaffiltext{27}{The Graduate University for Advanced Studies (SOKENDAI), Shonan International Village, Hayama-cho, Miura-gun, Kanagawa 240-0193, Japan}
\altaffiltext{28}{Department of Astronomy and Astrophysics, University of Toronto, 50 St. George Street M5S 3H4, Toronto Ontario, Canada}

\begin{abstract}
To reveal the structures of a transition disk around a young stellar object in Lupus, \object{Sz\ 91}, we have performed aperture synthesis 345 GHz continuum and CO(3--2) observations with the Submillimeter Array ($\sim1\arcsec$--3$\arcsec$ resolution), and high-resolution imaging of polarized intensity at the $K_s$-band by using the HiCIAO instrument on the Subaru Telescope ($0\farcs25$ resolution).
Our observations successfully resolved the inner and outer radii of the dust disk to be 65 and 170 AU, respectively, which indicates that \object{Sz\ 91} is a transition disk source with one of the largest known inner holes.
The model fitting analysis of the spectral energy distribution reveals an H$_2$ mass of $2.4\times10^{-3}$ $M_\sun$ in the cold ($T<$30 K) outer part at $65<r<170$ AU by assuming a canonical gas-to-dust mass ratio of 100, although a small amount ($>3\times10^{-9}$ $M_\sun$) of hot ($T\sim$180 K) dust possibly remains inside the inner hole of the disk.
The structure of the hot component could be interpreted as either an unresolved self-luminous companion body (not directly detected in our observations) or a narrow ring inside the inner hole.
Significant CO(3--2) emission with a velocity gradient along the major axis of the dust disk is concentrated on the \object{Sz\ 91} position, suggesting a rotating gas disk with a radius of 420 AU.
The \object{Sz 91} disk is possibly a rare disk in an evolutionary stage immediately after the formation of protoplanets because of the large inner hole and the lower disk mass than other transition disks studied thus far.
\end{abstract}


\section{Introduction}
Nearly all newly-formed stars are surrounded by disks of gas and dust, which provide the building blocks of planets \citep{bib:evans2009}.
Thus, studying the structure and evolution of these protoplanetary disks provides information about how and when planets form.
Initially, the disks are optically-thick, producing broadband infrared (IR) emission well in excess of the stellar photosphere.
The star+disk system has a rather flat IR spectral energy distribution (SED) slope \citep[e.g.,][]{bib:lada1984,bib:lada1987}.
By $\sim5$ Myr, nearly all stars lack evidence of warm circumstellar dust and accretion onto the star \citep[e.g.,][]{bib:fedele2010,bib:currie2011}.
This implies that most protoplanetary disks around solar/subsolar-mass stars have disappeared by this time, and the disk material has been accreted/dispersed because of processes such as photoevaporation or the disk material has been incorporated into planets.\par

Transition disks bridge the gap between these endpoints of disk evolution because their excess emission at some IR wavelengths is intermediate between those of an optically thick disk and a star lacking a disk \citep[e.g.,][]{bib:strom1989}.
Although the term ``transition disk'' includes a diverse set of morphologies, a major subset of these objects include those with near/mid IR deficits (relative to an optically-thick disk) but optically-thick emission at other, typically longer wavelengths, which is indicative of inner holes or gaps \citep[e.g.,][]{bib:calvet2002}.
IR SED modeling implies that the hole or gap sizes for these objects typically range between $\sim1$ and $\sim50$ AU \citep[e.g.,][]{bib:espaillat2010,bib:merin2010} and encloses the planet-forming region in our own solar system.
SED modeling has identified numerous other transition disks in nearby star forming regions \citep[e.g.,][]{bib:cieza2010,bib:merin2010,bib:currie2011,bib:romero2012,bib:cieza2012b}.
Because formation of gas giant planets creates holes or gaps in disks, transition disks with SEDs consistent with these features may be excellent laboratories for studying planet formation.

Although SED modeling can provide indirect evidence of an inner hole or gap, high-resolution observations with long baseline (sub-)millimeter interferometers and large aperture IR telescopes provide direct evidence of their existence \citep[e.g,][]{bib:brown2009,bib:hughes2009,bib:andrews2011,bib:isella2012,bib:cieza2012a,bib:mathews2012,bib:thalmann2010,bib:hashimoto2012,bib:mayama2012}.
Since millimeter and micron sized dust grains contribute to the majority of millimeter and NIR emissions, it is important to compare the detailed spatial distributions of these tracers for understanding the physics of the transition disks.\par

\object{Sz 91} is an M0.5 young star surrounded by a transition disk located in the Lupus III molecular cloud \citep[$d=200$ pc;][]{bib:comeron2008}.
The stellar position, ($\alpha_\mathrm{J2000}$, $\delta_\mathrm{J2000}$), is (16$^\mathrm{h}$ 07$^\mathrm{m}$ 11$\fs$6, $-39\arcdeg$ 03$\arcmin$ 47$\farcs$2).
The stellar mass and age have been estimated to be 0.49 $M_\sun$ and 5 Myr, respectively \citep{bib:hughes1994}.
The H$\alpha$ line width at 10 \% of the peak has been measured to be 283 km s$^{-1}$, suggesting accretion with $10^{-10}$ $M_\sun$ yr$^{-1}$ \citep{bib:romero2012}.
\object{Sz 91} is classified as a wide binary with a separation of 9\arcsec \citep{bib:melo2003}, corresponding to 1800 AU.
However, such a wide binary system is unlikely to disturb the disk evolution of a host star \citep{bib:kraus2012,bib:harris2012}.
Moreover the difference between their proper motions implies that these sources are not co-moving \citep{bib:roeser2010}.
We therefore treat the \object{Sz 91} disk as a circumstellar disk around a single star system throughout this paper.\par

The SED of \object{Sz 91} is characteristic of the transition disk.
The following remarkable features of this source can be identified in the SED: no significant IR excess with a spectral index from $K_\mathrm{s}$ to 24 $\micron$ of $-2$ (i.e., class\ III in the IR categorization), presence of a large dip of approximately 20 $\micron$, and very steep flux density rising between 24 and 70 $\micron$ \citep{bib:evans2009,bib:romero2012}.
\citet{bib:romero2012} categorized \object{Sz 91} as a giant planet-forming disk on the basis of the following features: very steep increase in flux density at 24 $\micron$, which indicates a sharp edge at the inner radius of the disk; a clear sign of mass accretion onto the central star; a relatively massive disk ($\sim5\times10^{-3}$ $M_\sun$).\par

Although most transition disks with inner holes have flux deficits restricted to 1--10 $\micron$ and optically thick emission at longer wavelengths, \object{Sz 91} differs from them because its flux deficit extends to significantly longer wavelengths (24 $\micron$), and it has strong far-IR to millimeter emission.
The total flux density of the 870 $\micron$ (345 GHz) continuum emission of \object{Sz 91} has been estimated to be 34.5$\pm$2.9 mJy \citep{bib:romero2012}.
Although the 1.3 mm continuum emission had not been detected previously \citep{bib:nurnberger1997}, the recent wide field imaging survey with the AzTEC receiver on the Atacama Submillimeter Telescope Experiment (ASTE) has clearly detected a 1.1 mm flux density of 27.2$\pm$6.0 mJy \citep{bib:kawabe2014}.
These submillimeter flux densities imply that the \object{Sz 91}'s disk is substantially massive with respect to those of other class\ III sources in nearby star forming regions \citep{bib:andrews2005,bib:andrews2007}; the flux densities roughly correspond to 2--5$\times$10$^{-3}$ $M_\sun$ if the canonical gas-to-dust mass ratio of 100 is assumed.
Such an object having significant submillimeter flux density, but no NIR excess appears to be at the transition phase from class\ II to III and is quite rare \citep[$\sim$5 \%][]{bib:andrews2005,bib:andrews2007}.
Although a recent high-resolution imaging survey for transition disk objects \citep{bib:andrews2011} has successfully revealed a large inner hole in the disk up to a radius of $\sim70$ AU, there is selection bias for the most massive transition disks.
Because the submillimeter flux density of \object{Sz 91} is the lowest among the transition disk objects studied thus far, considering the distance to the source \citep[Fig.10 in][]{bib:andrews2011}, \object{Sz 91} is a crucial target for investigating the disk evolution from the perspective of a variety of submillimeter flux densities, i.e., the disk mass.
However, the disk structure has not been resolved thus far, and high-resolution imaging observations are urgently required.\par

In this paper, we present the first sub-arcsecond resolution images toward \object{Sz 91} at submillimeter and near infrared wavelengths. 
In section \ref{sec:observation}, observational parameters and calibration for the obtained data are described.
In section \ref{sec:result}, high-resolution images are shown.
In section \ref{sec:discussion}, the detailed disk structure is discussed on the basis of model fitting to the SED and the CO(3--2) line profile. 
The unique nature of \object{Sz 91} is determined from a comparison with other transition disks.
Finally in section \ref{sec:summary}, the key results of this study are summarized.\par

\section{Observations and Data Reduction}\label{sec:observation}

\subsection{High-resolution Imaging of the 345 GHz Continuum and CO(3--2) Emission}
Interferometric observations of the 345 GHz continuum and CO(3--2) emission toward \object{Sz 91} were conducted in 2010 with the Submillimeter Array \citep[SMA;][]{bib:ho2004}, which comprises eight 6 m antennas located atop Mauna Kea in Hawaii.
The observations were performed in the compact configuration with baselines ranging from 6 to 70 m.
For the continuum emission, higher resolution data were also obtained in 2010 in the very extended configuration with baselines up to $\sim500$ m.
The field of view of SMA was approximately $30\arcsec$ in full width at half maximum (FWHM).
The emission was detected using double side-band superconductor-insulator-superconductor receivers with a local oscillator frequency of $\nu_\mathrm{LO}=340.755$ GHz.
Both the upper and lower sidebands data were used for the continuum, resulting in an 8 GHz bandwidth in total.
The channel spacing of the CO(3--2) line was set to 0.8125 MHz.\par

The amplitude and phase of the array system were calibrated by observations of quasars, J1604$-$446, J1626$-$298, and J1517$-$243, in a cycle with 12.5 min on a target and 6 or 8 min on a quasar.
The response across the observed passbands was determined by 60 min observations of quasar, \mbox{3C 273}.
Absolute flux calibration was achieved by observing Titan at the beginning and end of each night.

We created the continuum images by combining the very extended and compact configuration data, whereas the CO maps were created from the compact configuration data only.
The UV data were edited and calibrated using MIR, an IDL-based software package.
We used Astronomical Image Processing System (AIPS) for imaging procedures, including deconvolution by the CLEAN algorithm and restoration with a synthesized beam.
Both continuum and line emission maps were created with natural weighting in the visibility plane, producing synthesized beam FWHMs of $1\farcs4\times1\farcs0$ at a position angle (PA) of $-9\fdg8$ and $4\farcs6\times1\farcs8$ at a PA of $-4\fdg0$, respectively.\par

\subsection{High-resolution Polarized Intensity Imaging at $K_s$ band with the Subaru Telescope}
$K_s$-band (2.15 $\micron$) linear polarized intensity (PI) images of \object{Sz 91} were obtained with the high-contrast imaging instrument \citep[HiCIAO;][]{bib:tamura2006} combined with dual-beam polarimetry from the 8.2 m Subaru Telescope in May 2012.
The observations were conducted under the program SEEDS \citep[Strategic Explorations of Exoplanets and Disks with Subaru; ][]{bib:tamura2009}.
The adaptive optics system \citep[AO188:][]{bib:hayano2004} provided a limited diffraction and mostly stable stellar point spread function (PSF) with a FWHM of $0\farcs06$ in the $K_s$ band.
Polarization differential imaging (PDI) is a powerful technique used to reveal the structure of a dusty disk in very close proximity to a star \citep[e.g.,][]{bib:thalmann2010,bib:hashimoto2011,bib:hashimoto2012,bib:tanii2012,bib:muto2012,bib:kusakabe2012,bib:mayama2012}.
We employed the PDI mode, combined with the angular differential imaging mode, in which the field of view and the pixel scale were $10\arcsec \times 20\arcsec$ and 9.5 mas pixel$^{-1}$, respectively.
Half-wave plates were placed at the four angular positions of $0\degr$, $45\degr$, $22\fdg5$, and $67\fdg5$ in sequence with a 30 sec exposure per wave plate position.
Image Reduction and Analysis Facility (IRAF\footnote{IRAF is distributed by National Optical Astronomy Observatory, which is operated by the Association of Universities for Research in Astronomy, Inc., under cooperative agreement with the National Science Foundation.}) software was used for all data reduction as following the methods of \citet{bib:hashimoto2011} and \citet{bib:tanii2012}, and the Stokes {\it Q} and {\it U} parameter images were created.
We calculated the PI as $\sqrt{Q^2+U^2}$ and the polarization vector angle as $0.5\times \arctan(U/Q)$ with the 3.4.0 version of the Common Astronomy Software Applications package.
The final PI image was created from the smoothed {\it Q} and {\it U} images by a Gaussian function; thus, the effective resolution of the image became $\sim0\farcs25$.

\section{Results}\label{sec:result}

\subsection{345 GHz Continuum Emission Map}\label{subsec:sma_cont}
\begin{figure}
	\epsscale{1.0}
	\plotone{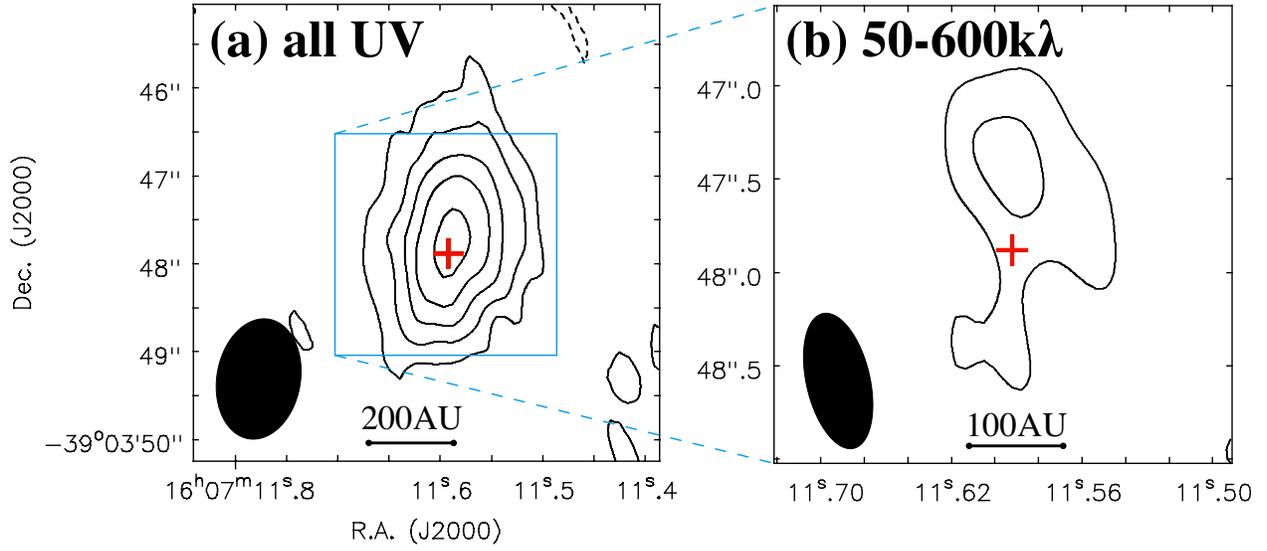}
	\caption{CLEANed maps of the 345 GHz continuum emission created by all UV data (a) and 50--600 k$\lambda$ (b). The contours begin at $\pm2\sigma$ with the intervals of 2$\sigma$, where 1$\sigma=$2.1 and 2.8 mJy beam$^{-1}$ in (a) and (b), respectively. The red cross indicates the stellar position in each panel. The synthesized beam sizes in half power beamwidth (HPBW) are shown at the bottom-left corner in each panel: $1\farcs4\times1\farcs0$ with a position angle (PA)=$-9\fdg8$ and $0\farcs7\times0\farcs3$ with a PA=$13\fdg0$ for (a) and (b), respectively. The blue box in the left panel indicates the area of the right panel.}\label{fig:sma_cont}
\end{figure}

We created a 345 GHz continuum emission map of \object{Sz 91} with all UV data as shown in Figure \ref{fig:sma_cont}(a).
The map shows a strong continuum emission peak at the stellar position.
The total flux density is measured to be 32.1$\pm$3.6 mJy, which agrees well with the flux density of 34.5$\pm$2.9 mJy, which is the previous 345 GHz measurement with the 12 m single dish APEX telescope obtained by \citet{bib:romero2012}.
The 345 GHz emission is resolved by the $1\farcs4\times1\farcs0$ beam at a PA of $-9\fdg8$.
The beam-deconvolved size of the emission is measured to be (1\farcs7$\pm$0\farcs1)$\times$(0\farcs7$\pm$0\farcs2) with a PA of 169\fdg0$\pm$3\fdg0 from a 2-D Gaussian fitting, corresponding to (340$\pm$20)$\times$(140$\pm$40) AU at a distance of 200 pc.
This size is comparable to that of a typical protoplanetary disk.
The 345 GHz emission is probably thermal dust emission from the disk around \object{Sz 91} because the flux densities from 350 to 1100 $\micron$ monotonically decrease with a spectral index of $\sim2$ (Figure \ref{fig:sed}), which is comparable to other T Tauri stars \citep{bib:andrews2005,bib:andrews2007}.\par

Moreover, we created a higher-resolution 345 GHz image using the UV data over the UV range from 50 to 600 k$\lambda$ as shown in Figure \ref{fig:sma_cont}(b).
The emission peak at the 4.5$\sigma$ level can be found to the north of the star.
The peak position is measured to be ($\alpha_\mathrm{J2000}=$16$^\mathrm{h}$ 07$^\mathrm{m}$ 11$\fs$6, $\delta_\mathrm{J2000}=$ $-39\arcdeg$ 03$\arcmin$ 47$\farcs$8) and is shifted from the stellar position by $0\farcs43$ toward the direction at a PA of $1\fdg4$.
The lack of a peak at the stellar position implies that the inner part of the dust disk is depleted or cleared as expected from the SED.
Therefore, the continuum emission in Figure\ \ref{fig:sma_cont}(b) probably originated from the innermost part of the dust disk.
Assuming that the peak position in figure\ \ref{fig:sma_cont}(b) represents the inner edge of the dust disk, the inner radius of the Sz\ 91 transition disk is estimated to be 86$\pm$25 AU.
The total flux density of this compact component is measured to be 13.2$\pm$3.1 mJy, which is 40 \% of the total flux density in Figure \ref{fig:sma_cont}(a).\par

\subsection{CO(3--2) Molecular Line Emission Maps}\label{subsec:sma_co}
\begin{figure}
	\epsscale{1.0}
	\plotone{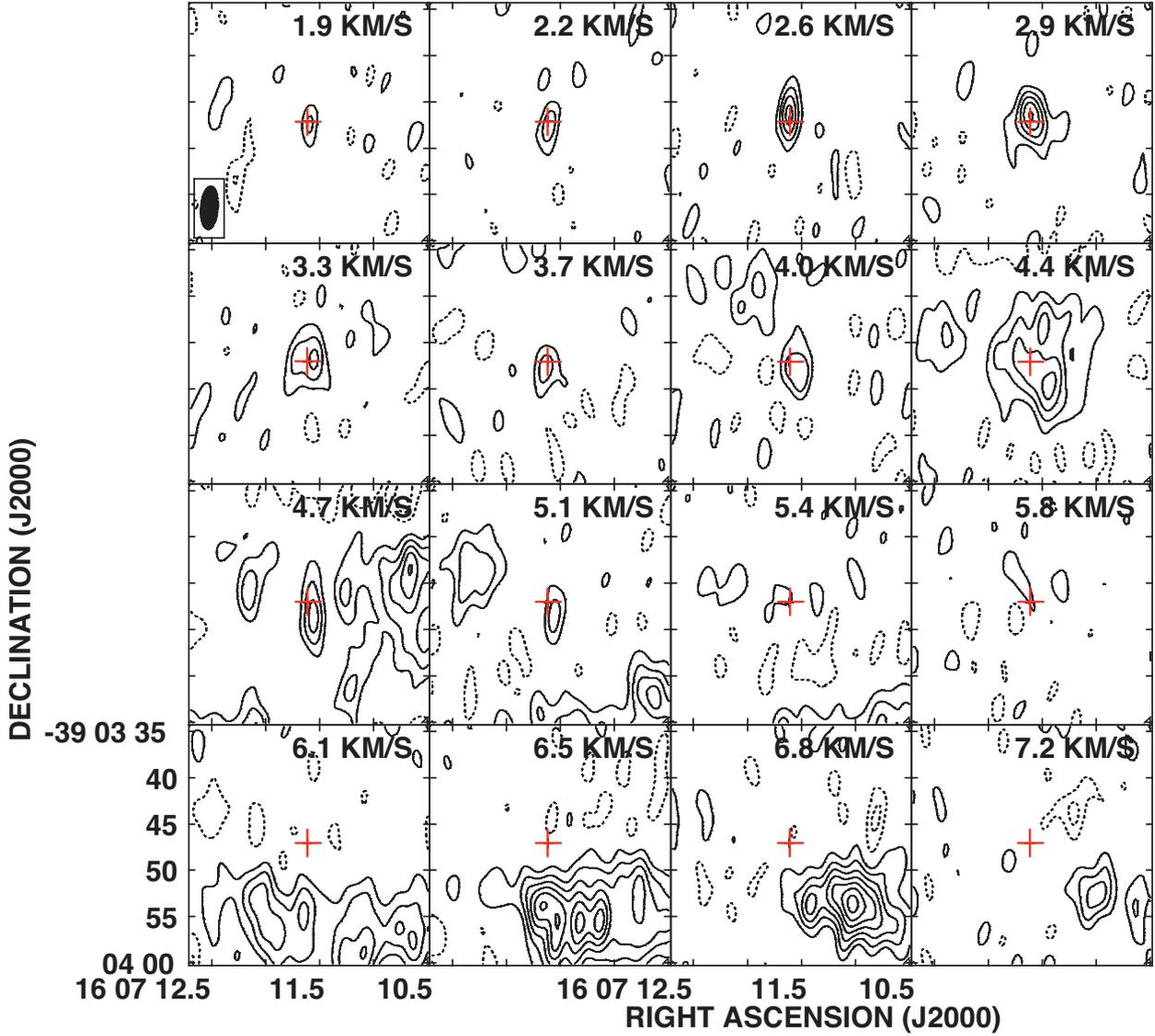}
	\caption{CO(3--2) velocity channel maps with 0.35 km s$^{-1}$ resolution obtained by SMA. The central LSR velocity of each panel is shown at the top in km s$^{-1}$. The contours start at $\pm2\sigma$ with intervals of $2\sigma$, where $1\sigma=289$ mJy beam$^{-1}$. The ellipse at the bottom left in the top left panel shows the synthesized beam size of $4\farcs6\times1\farcs8$ with a PA of $-4\fdg0$ in HPBW.}\label{fig:sma_co_ch}
\end{figure}

Figure \ref{fig:sma_co_ch} shows velocity channel maps of the CO(3--2) line.
The significant CO(3--2) emission is detected in the velocity range from 1.9 to 7.2 km s$^{-1}$ in $V_\mathrm{LSR}$.
It is clear that the emission at $V_\mathrm{LSR}=$1.9--5.8 km s$^{-1}$ is concentrated at the stellar position.
In addition, a velocity gradient appears to be present along the north-south direction in this velocity range: the emission is located mainly on the north side of the star at $V_\mathrm{LSR}=$2.6--2.9 km s$^{-1}$ and at the south from 4.0 to 5.1 km s$^{-1}$.
Notably, the central velocity range from 3.3 to 4.0 km s$^{-1}$ includes the radial velocity of \mbox{Sz 91} \citep[$V_\mathrm{LSR}=$3.87 km s$^{-1}$, converted from $V_\mathrm{helio}=-1.57$ km s$^{-1}$;][]{bib:melo2003}.
In the range of 4.4--5.4 km s$^{-1}$, we detected spatially extended emission near the star (4.4 and 4.7 km s$^{-1}$) and at the south side of the field of view (4.7--5.4 km s$^{-1}$).
The origin of the extended components is most probably an ambient cloud because the systemic velocity and the velocity width of the main cloud condensation of Lupus\ III have been measured to be 4.1 and 1.2 km s$^{-1}$, respectively \citep{bib:hara1999}.
We also detected the extended emission in the 6.1--7.2 km s$^{-1}$ channels.
These components are shifted by $\gtrsim3\arcsec$ toward the south-west direction from the star and also possibly originate from the spatially extended ambient cloud.
Hereafter, we focus on compact components toward the star that should have originated from the gas disk; we do not discuss the extended components in this paper.\par

\begin{figure}
	\epsscale{1.0}
	\plotone{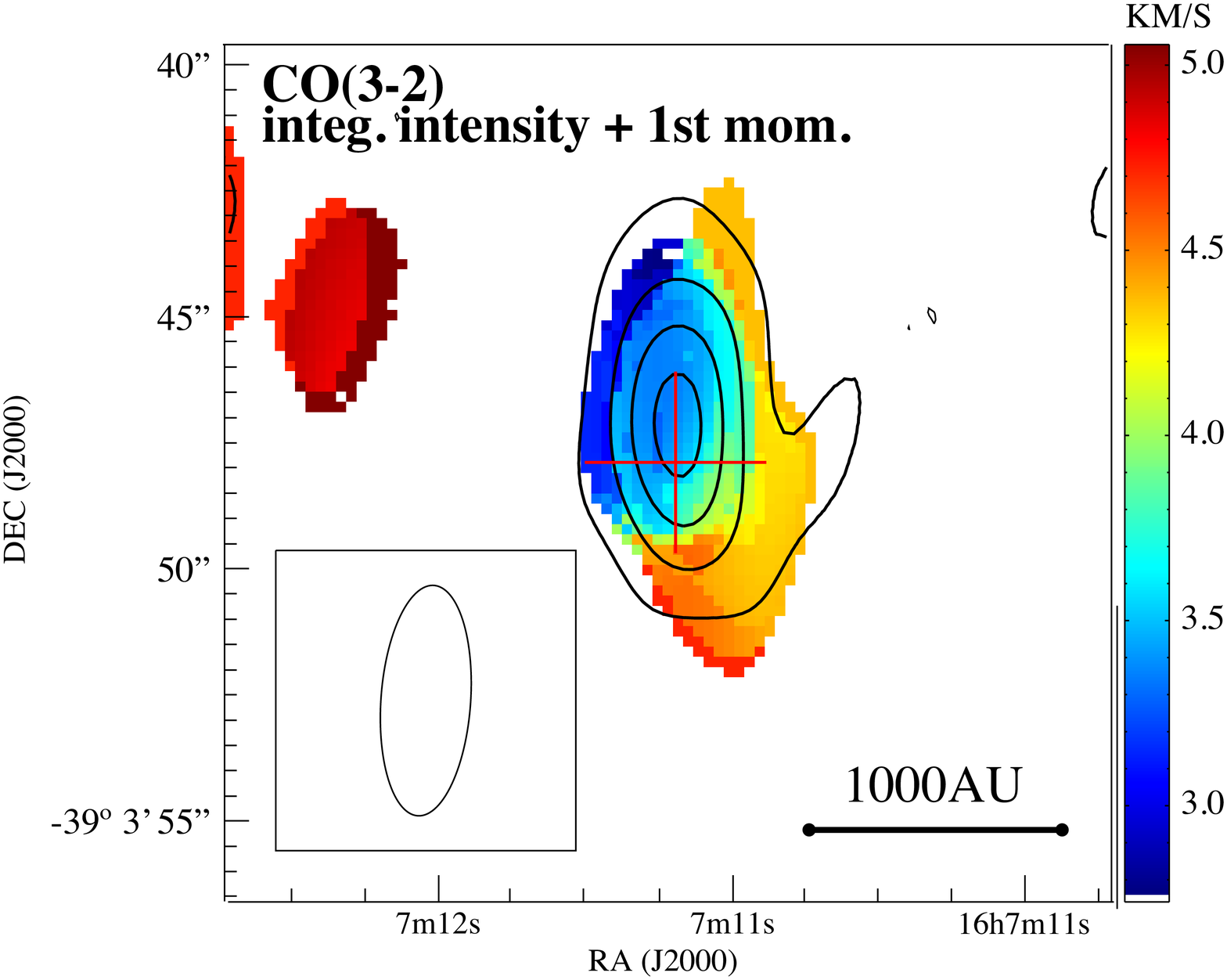}
	\caption{Total integrated intensity map of the CO(3--2) emission over a $V_\mathrm{LSR}$ range of 1.5--6.1 km s$^{-1}$ (contours) superimposed on the intensity-weighted first moment map (color). The contours start at $\pm3\sigma$ with intervals of 3$\sigma$, where 1$\sigma=$450 mJy km s$^{-1}$ beam$^{-1}$. The color bar on the right-hand side of the panel shows the centroid velocity in m s$^{-1}$. The red cross indicates the stellar position. The synthesized beam size in HPBW is shown at the bottom left corner: $4\farcs6\times1\farcs8$ with PA$=-4\fdg0$.}\label{fig:sma_co_int}
\end{figure}

To clearly see the gas disk emission and its velocity gradient, we created total integrated intensity (1.5--6.1 km s$^{-1}$) and first moment maps of the CO(3--2) line, as shown in Figure \ref{fig:sma_co_int}.
From these maps, we successfully detected the centrally concentrated emission and the velocity gradient roughly along the north-south direction, suggestive of the rotating gas disk.
The total integrated intensity of CO(3--2) in the area above the 3$\sigma$ noise level is 8.55 Jy km s$^{-1}$.
The emission peak is shifted from the stellar position by $\sim1\arcsec$ toward the north direction, possibly because the red-shifted emission is partially resolved-out owing to the contamination by the ambient cloud.\par

\subsection{Polarized Intensity Image at $K_s$ band}\label{subsec:kband}
\begin{figure}
 \epsscale{1.0}
 \plottwo{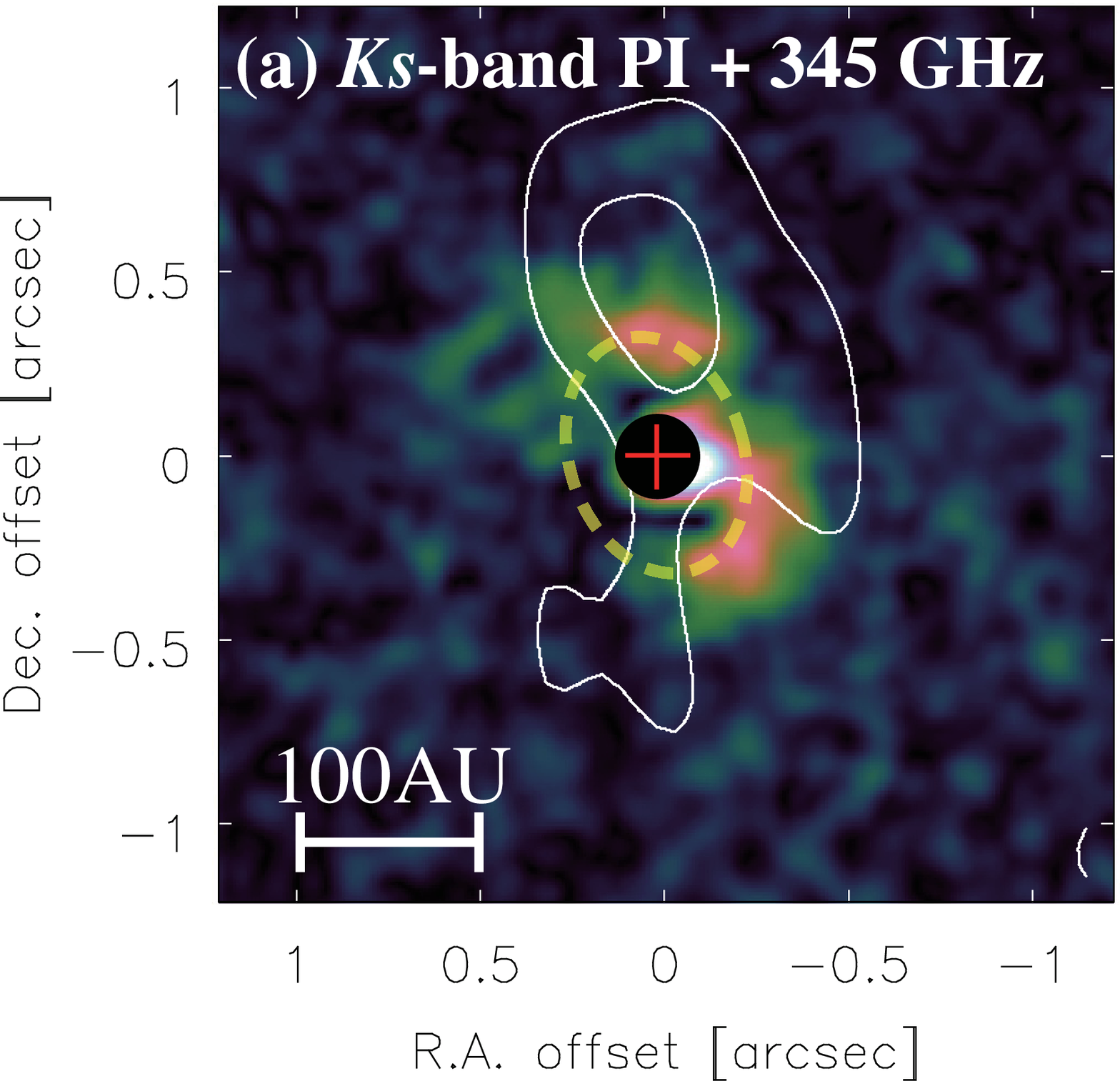}{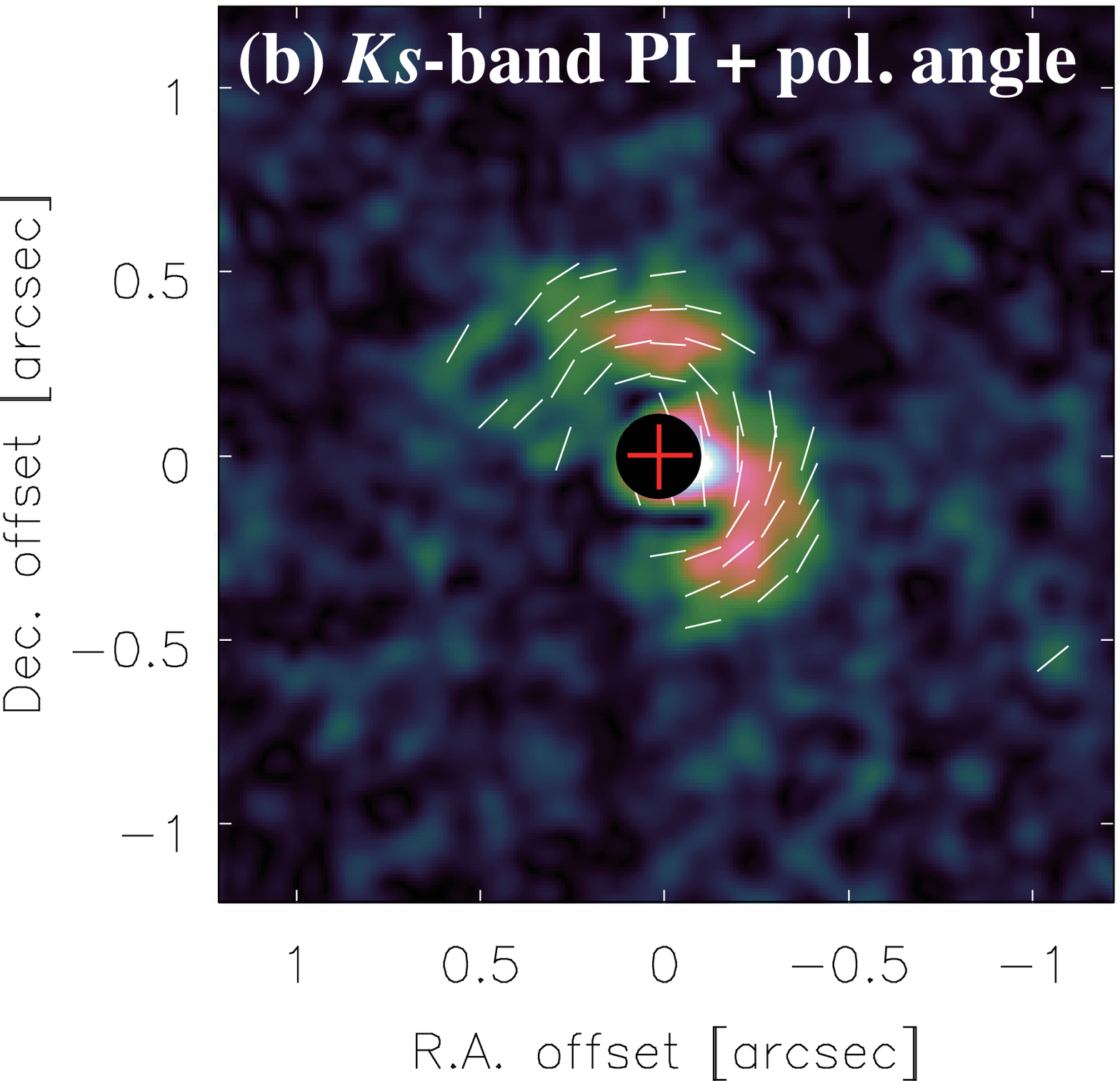}
 \caption{(a) Polarized intensity (PI) image obtained with the Subaru telescope in the $K_s$ band (color) where the 345 GHz continuum emission map (contour) in Figure \ref{fig:sma_cont}(b) is superposed. The stellar position is shown by the red cross. The best-fit ellipse is indicated by the yellow dashed line. The inner area of $0\farcs25$ diameter (filled circle) is photometrically unreliable owing to the point spread function (PSF) subtraction process and is masked. (b) Polarization vector angle map superimposed on the PI image in the left panel. The vector directions indicate angles of polarization. The vector's lengths are arbitrary.}\label{fig:kband}
\end{figure}

The high-resolution PI and polarization vector angle maps at the $K_s$-band are shown in Figures\ \ref{fig:kband}(a) and (b), respectively.
The presented images are smoothed and the effective resolution of the images is $\sim0\farcs25$.
A crescent-like emission region was detected around \object{Sz 91}, which is elongated from the south to the north via the west side of the star.
The polarization angles are nearly perpendicular to the radial directions from the central star, indicating that the $K_s$-band emission probably originated from the scattered light at the inner part of the dust disk.
The crescent-like emission suggests that this part is the near side of the disk if we assume that forward scattering is dominant, as is the case in Mie scattering.
Substantial emission dips appear at the north and south sides of the software mask, possibly due to the inner hole structure of the dust disk.
The PSF subtraction process is a primary factor to cause an artificial systematic error near the star, and the error becomes a systematic emission via the square root operation of stokes Q and U.
The symmetric distribution of the dips indicates the existence of an inner hole in the dust disk.\par

We also detected a bright region near the western edge of the software mask, i.e., a photometrically reliable emission.
However, the origin of this emission remains unclear.
Because the PSF shows a substantially symmetric distribution, it is unlikely that the bright emission would appear at the only one-side.
Because it extends to the edge of the software mask, higher resolution and higher contrast observations close to the star are required to determine the origin of the bright emission.\par

The existence of the inner hole structure expected from the PI image strongly supports the results of the submm images.
The higher-resolution 345 GHz image is overplotted in Figure \ref{fig:kband}(a).
It is clear that the PI is distributed at the innermost part of the submm emission.
For example, along PA=0$\degr$, the 345 GHz emission peaks at $0\farcs45$ from the star whereas the PI peaks at $0\farcs32$.
These facts suggest the existence of the hole structure at the inner part of the disk and that the PI originates from the innermost part of the transition disk, i.e., the inner edge.\par

To quantify the shape of the inner edge of the dust disk, we fit an ellipse to the crescent-like emission.
Table \ref{tab:ellipse} lists the parameters of the best-fit ellipse, and the fitted ellipse is shown in Figure \ref{fig:kband}(a) by the yellow dotted line.
The center position of the ellipse coincides with the stellar position.
The PA of the ellipse, i.e., the direction of the disk major axis, is along at $17\fdg5\pm17\fdg7$, which is consistent with the PA of the emission dips in the crescent-like emission.
The PI at the disk minor axis (PA$\sim120\degr$) appears to be lower than that at the major axis.
This result is probably due to lower polarization degree along the minor axis where the scattering angle at the disk surface is deviated from 90$\degr$ \citep{bib:mccabe2002}.\par

\begin{deluxetable}{lcc}
\tabletypesize{\small}
\tablecaption{Parameters of the best-fit ellipse in Figure \ref{fig:kband}}
\tablewidth{0pt}
\tablehead{
\colhead{Parameter} &
\colhead{Best-fit value} &
\colhead{Error}
}
\startdata
$\Delta$R.A.\tablenotemark{1} ($\arcsec$) & 0.03 & 0.04 \\
$\Delta$Dec.\tablenotemark{1} ($\arcsec$) & 0.03 & 0.02 \\
Major axis ($\arcsec$) & 0.33 & 0.02 \\
Minor axis ($\arcsec$) & 0.25 & 0.08 \\
Position angle ($\degr$) & 17.5 & 17.7
\enddata
\tablenotetext{1}{Offset from the stellar position.}
\label{tab:ellipse}
\end{deluxetable}

\section{Discussion}\label{sec:discussion}

\subsection{Spatially Resolved Disk Structure}\label{sec:disk_structure}
Both the high-resolution submm and NIR images clearly resolved the disk structure around \object{Sz 91}.
In this section, we estimate the inner radius, outer radius, and inclination angle of the disk directly from our images.
These parameters are adopted in the model calculations in sections \ref{sec:sedcalc} and \ref{sec:linecalc}.\par

The inner radius of the dust disk can be estimated from the large inner hole revealed in the $K_s$-band PI image.
Assuming that the $K_s$-band emission is the scattered light at the inner edge of the disk, the best-fit ellipse in the $K_s$-band image shows the inner radius to be 65$\pm$4 AU.
There are two possibilities to explain the scattered light.
One is forward scattering at the surface of the inner part of the disk, and the other is scattering at the inner wall of the disk such as that in the case of \mbox{LkCa\ 15} \citep{bib:thalmann2010}.
We can not judge which case is more likely to explain the NIR emission from the Sz 91 disk.
However, in either case, it probably originates from the innermost part of the disk, and our estimation of the inner radius seems to be reasonable.
The high-resolution 345 GHz image also provides a rough estimate of the disk inner radius.
Although the 345 GHz image in Figure \ref{fig:sma_cont}(b) does not clearly show the presence of the inner hole, it is clear that the 345 GHz flux density becomes weak toward the central star, which is suggestive of dust depletion near the star.
The angular distance between the 345 GHz emission peak and the central star can be interpreted as the disk inner radius of $0\farcs43$ ($=86$ AU), which is roughly consistent with the above estimate of 65 AU.
In the following discussion, we adopt the disk radius $R_\mathrm{in}$ of 65 AU.\par

The outer radius of the dust disk can be estimated from the beam-deconvolved size of the full-UV 345 GHz map shown in Figure\ \ref{fig:sma_cont}(a).
Assuming a geometrically thin disk, the beam-deconvolved size of (1\farcs7$\pm$0\farcs1)$\times$(0\farcs7$\pm$0\farcs2) corresponds to the outer radius $R_\mathrm{out}$ of 170$\pm$20 AU at the distance of 200 pc.
The gas disk is also resolved in CO(3--2) (Figure\ \ref{fig:sma_co_int}).
The beam-deconvolved size is measured to be (3\farcs1$\pm$0\farcs3)$\times$(1\farcs6$\pm$0\farcs7) at a PA of 15$\pm$11\degr from a 2-D Gaussian fitting, corresponding to $R_\mathrm{out}=$310$\pm$60 AU.
However, the red-shifted gas is probably contaminated by the ambient cloud gas, and the center position of the beam-deconvolved disk is shifted by $0\farcs6$ from the star to the north.
Thus we consider the largest distance of $2\farcs1$ from the stellar position as the outer radius of the gas disk: $R_\mathrm{out}=$420 AU.
Because the actual intensity distribution of each emission is expected to have a power-law form, these outer radii derived from the Gaussian fitting should be regarded as lower limits.\par

The outer radius of the gas disk is much larger than that of the dust disk.
The discrepancy in radial extension between the dust and gas emissions is frequently observed for other protoplanetary disks, which can be explained by an exponential decrement in the surface density \citep[e.g.,][]{bib:hughes2008,bib:panic2009,bib:isella2010} or by the radial drift of large grains \citep{bib:andrews2012}.
In addition, the possible north-south asymmetry in the dust continuum emission, suggested in Figure \ref{fig:sma_cont}(b), could have resulted from the azimuthal drift of large grains \citep{bib:birnstiel2013,bib:vandermarel2013} associated perhaps with a perturbing body orbiting within the central hole of this disk.\par

The disk inclination angle can be derived from the ratio of the major to minor axis lengths of the best-fit ellipse in Figure \ref{fig:kband}(a): 40$\pm$15$\degr$ from face-on.
Moreover, the beam-deconvolved size of the 345 GHz continuum emission in Figure \ref{fig:sma_cont}(a) provides an inclination angle of 66$\pm$5$\degr$.
The latter value of 66$\degr$, however, is possibly influenced by the disk flaring because the distribution of the 345 GHz emission is biased to the outer part of the disk.
Thus, we adopt the $K_s$-band value of 40$\degr$ as the disk inclination angle in this study.
In fact, for the CO(3--2) spectrum, the model calculation in section 4 prefers the smaller value of 40$\degr$.\par

\subsection{Disk Parameters Derived from SED fitting}\label{sec:sedcalc}
\begin{figure}
 \epsscale{1.0}
 \plotone{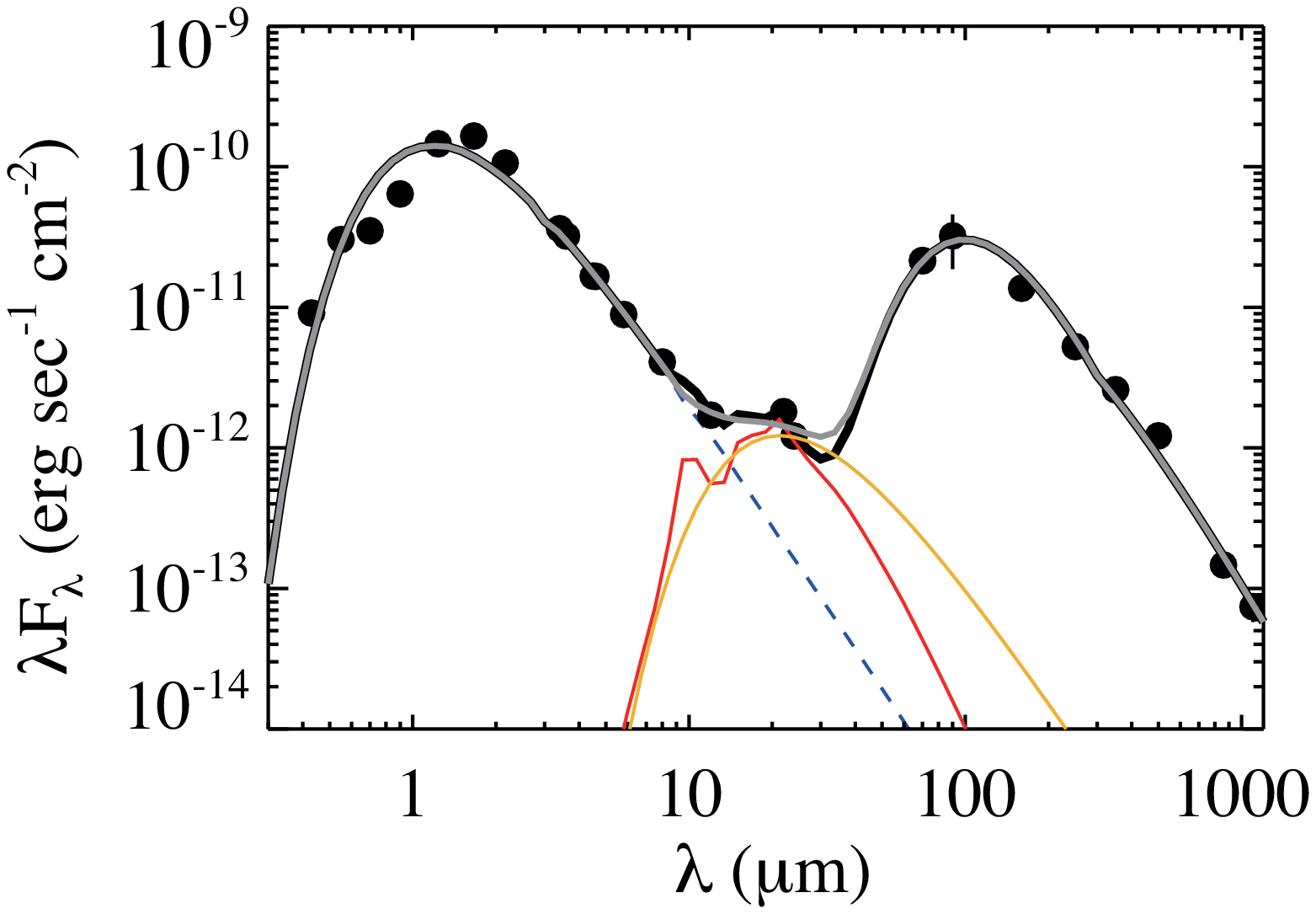}
 \caption{Spectral energy distribution of \mbox{Sz 91}. The filled circles indicate observed flux densities. All data except the 345 GHz flux density were compiled from previous studies, including the NOMAD catalog \citep{bib:zacharias2005}, the {\it 2MASS} point source catalog \citep{bib:cutri2003}, the {\it Spitzer} IRAC and MIPS photometry \citep{bib:evans2009}, the {\it WISE} all-sky data release \citep{bib:cutri2012}, the {\it AKARI} FIS all-sky survey point source catalog \citep{bib:yamamura2010}, the {\it IRAS} point source reject catalog \citep{bib:iras2007}, the AzTEC/ASTE 1100 $\micron$ photometry \citep{bib:kawabe2014}, and the {\it Herschel} PACS/SPIRE FIR flux densities which we derived from the HSA Science archive data. The best-fit SED and the contribution of the hot component in the optically thin case are shown by the black and red lines, respectively, and by the gray and orange lines in the optically thick case. The blue dashed line indicates the stellar contribution.}\label{fig:sed}
\end{figure}

To deduce other parameters of the dust disk, we performed least-square fitting to the SED on the basis of the power-law disk model.
Figure \ref{fig:sed} shows the SED of \object{Sz 91} that includes our 345 GHz flux density and a recent data set by {\it WISE}.
As \citet{bib:romero2012} reported, the SED is characterized by large dip around 20 $\micron$, sharp rise from 20 to 70 $\micron$, and significant (sub-)millimeter emission.\par

We introduced the two components of a cold disk and a small amount of hot dust inside the disk as follows.
First, we applied the model for the cold disk, which comprised the usual power-law disk and blackbody stellar emissions, to all of the SED data except that at 10--30 $\micron$, because the model could not reproduce all the data points including the three points at 10--30 $\micron$.
The power-law disk model has surface density and temperature radial distributions of $\Sigma(r)$ and $T(r)$, respectively, in a power-law form \citep{bib:kitamura2002,bib:tsukagoshi2011}:
\begin{equation}
\Sigma(r)=\Sigma_\mathrm{in} \biggl( \frac{r}{R_\mathrm{in}} \biggr)^{-p}
\end{equation}
and
\begin{equation}
T(r)=T_\mathrm{in} \biggl( \frac{r}{R_\mathrm{in}} \biggr)^{-q} \mathrm{,}
\end{equation}
where $r$ is the radial distance, $\Sigma_\mathrm{in}$ and $T_\mathrm{in}$ are the surface density and temperature at the inner radius, $R_\mathrm{in}$, and $p$ and $q$ are power-law indexes.
The gas-to-dust ratio was assumed to be 100 and the extinction of the stellar light because of the interstellar dust was corrected by the $A_\mathrm{V}$ value toward the star and the dust mass absorption coefficient \citep[Figure 1 of][]{bib:adams1988}.
Table \ref{tab:sedfit1} lists the fixed parameters in the SED fitting.
The stellar mass, the effective temperature of the star, and the visual extinction toward the star are from \citet{bib:hughes1994}.
The inner and outer radii and inclination angle of the disk were determined in section \ref{sec:disk_structure}.
We here adopted the values of 1.5 and 0.5 as $p$ and $q$, respectively, which are the same as those of the minimum mass solar nebular \citep{bib:hayashi1981}.
These values were selected because the SED fitting is known to be insensitive to the power-law index of $p$ and it is difficult to resolve the parameters for the temperature profile ($T$ and $q$) from only the longer wavelength data ($>30$ \micron).
Notably, the lower limit of the disk temperature was set to be 10 K, which is the typical temperature of the Lupus\ III cloud \citep{bib:vilas-boas2000}.
The stellar radius, $R_\ast$, $T_\mathrm{in}$, $\Sigma_\mathrm{in}$, and the power-law index of the dust mass opacity coefficient $\beta$ \citep[$\kappa_\nu=0.1\times(\nu/10^{12}\ \mathrm{Hz})^\beta$ cm$^2$ g$^{-1}$:][]{bib:beckwith1990} were treated as the free parameters in the SED fitting.\par

After the power-law disk model fitting, we introduced the additional {\it hot component} inside the disk to reproduce the observed flux densities at $\sim20$ $\micron$.
The presence of the inner hot component is supported by the sign of mass accretion derived from the H$\alpha$ emission line \citep{bib:romero2012}.
Because no information was available on the structure of this component, we simply assumed a gray body with a temperature of $T_\mathrm{c}$, a column density of $\Sigma_\mathrm{c}$, and a solid angle of $\Omega_\mathrm{c}$. 
Its flux density $S_\lambda$ at wavelength $\lambda$ is written by
\begin{equation}
S_\lambda = \frac{2 h c^2}{\lambda^5}  \frac{1}{\exp(\frac{hc}{\lambda k_\mathrm{B} T_\mathrm{c}})-1} (1-e^{-\kappa_\nu \Sigma_\mathrm{c}}) \times \Omega_\mathrm{c}\ \mathrm{,}
\end{equation}
where $h$ is the Plank constant, $c$ is the speed of light, and $k_\mathrm{B}$ is the Boltzmann constant.
We applied the same dust mass opacity coefficient, $\kappa_\nu$, as that in the cold disk.
The parameters of $T_\mathrm{c}$, $\Sigma_\mathrm{c}$, and $\Omega_\mathrm{c}$ were treated as free.
In the fitting, the upper limit of $\Omega_\mathrm{c}$ was set to be $\sim6.0\times10^{-12}$ str, corresponding to the solid angle of the inner hole seen in the NIR image.\par

The best-fit model SED of the cold disk reproduced the longer wavelength data ($>30$ \micron) effectively, as shown in Figure \ref{fig:sed}.
The best-fit parameters of the cold disk are summarized in Table \ref{tab:sedfit2}.
The $\beta$ value of 0.5$\pm$0.1 is significantly smaller than that in the diffuse interstellar medium of $\sim2$ \citep{bib:draine1984}, suggesting that the dust growth occurs in the disk \citep{bib:miyake1993}.
By adopting this $\beta$ and $\kappa_\nu$ introduced by \citet{bib:beckwith1990}, who assumed a 100:1 mass ratio between gas and dust, the disk mass is derived to be (2.4$\pm$0.8)$\times10^{-3}$ $M_\sun$.
This value is significantly higher than those of other class III sources in nearby star forming regions \citep{bib:andrews2005,bib:andrews2007}; however, the value is lower than those of most of the transition disks studied thus far \citep{bib:andrews2011}, even if we consider the difference in $\kappa_\nu$ by a factor of $\sim$1.7 at $\nu=340$ GHz between this study and the previous studies in which $\beta$ is set to be 1.\par

\begin{figure}
 \epsscale{1.0}
 \plotone{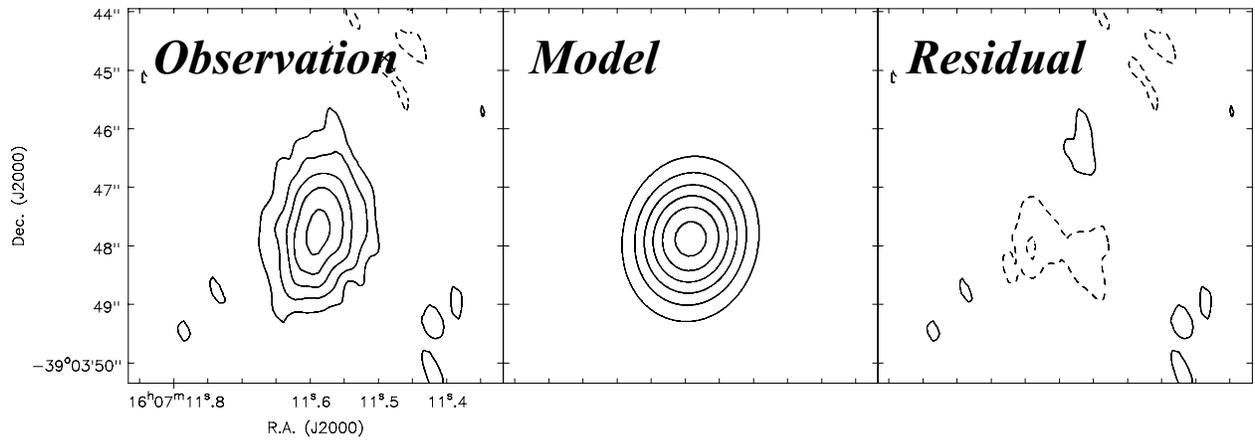}
 \caption{Comparison of the observed data with the best-fit model. The 345 GHz continuum image, model image, and residual image (data$-$model) are shown from left to right. The contour starts at $\pm2\sigma$, where 1$\sigma=$2.1 mJy beam$^{-1}$. The solid and dashed lines indicate positive and negative, respectively.}\label{fig:modelim}
\end{figure}

To verify the fitting result, we also created a continuum image for the best-fit model and compared it with the observations, as shown in Figure \ref{fig:modelim}.
The model image essentially agrees with the observations, but there remains a difference of at most $4\sigma$; the negative residual extends in the east-west direction, which may be due to the asymmetry of the disk.\par

We next attempted to reproduce the observed flux densities at 10--30 $\micron$ by adding the contribution from the hot component.
The best-fit parameters of $T_\mathrm{c}$, $\Sigma_\mathrm{c}$, and $\Omega_\mathrm{c}$ to reproduce all of the SED data were searched in the reduced $\chi^2$ maps by manually changing the initial values of $T_\mathrm{c}$, $\Sigma_\mathrm{c}$, and $\Omega_\mathrm{c}$ with 6000 runs.\par

From the calculations, we determined that there are two distinct regions in the $\Sigma_\mathrm{c}$-$\Omega_\mathrm{c}$ plane where the reduced $\chi^2$ takes its local minimum values, which indicates a strong coupling between $\Sigma_\mathrm{c}$ and $\Omega_\mathrm{c}$.
One is the region in which $\Sigma_\mathrm{c}<5\times10^{-5}$ g cm$^{-2}$ and $\Omega_\mathrm{c}>4\times10^{-13}$ str, and the emission of the hot component is substantially optically thin. 
In this region, the best-fit $T_\mathrm{c}$ typically converges to be 186 K and the best-fit $\Sigma_\mathrm{c}$ is inversely proportional to $\Omega_\mathrm{c}$.
The product of $\Sigma_\mathrm{c}\times\Omega_\mathrm{c}$ is a constant value, which provides the best-fit mass of $3\times10^{-9}$ $M_\sun$ for the hot component.
The other is the region in which $\Sigma_\mathrm{c}>30$ g cm$^{-2}$ and the emission is substantially opticall thick.
In this region, the best-fit $T_\mathrm{c}$ is typically 172 K and $\Sigma_\mathrm{c}$ is independent of $\Omega_\mathrm{c}$.
For the hot component, we can obtain the best-fit $\Omega_\mathrm{c}=1\times10^{-16}$ str and the lower limit mass of $6\times10^{-7}$ $M_\sun$.
Although the reduced $\chi^2$ values are slightly lower in the former case ($\Delta\chi^2=0.3$), the difference between the two regions is not significant.
We therefore conclude that there are two possible origins of the hot component: the optically thin gray body emission and the optically thick black-body emission.
However, the optically thin condition is the unlikely case because the mass of the hot component is too small given the mass accretion rate of $10^{-10}$ $M_\sun$ yr$^{-1}$; all the hot component would disappear in only 30 yrs.\par

Notably, the contribution of the hot component is mainly restricted by only three data points at MIR, from 12 to 25 $\micron$; the upper limit of the temperature is determined by the slight excess emission at 12 $\micron$ and the lower limit is limited by the decrement between 22 and 24 $\micron$.
A refined model will be required when more data at MIR and FIR are obtained by further observations.\par

\begin{deluxetable}{cc}
\tabletypesize{\small}
\tablecaption{Fixed parameters in the SED fitting}
\tablewidth{0pt}
\tablehead{
\colhead{Parameter} &
\colhead{Fixed Value}
}
\startdata
$M_\ast$ ($M_\sun$) & 0.49\tablenotemark{1} \\
$A_\mathrm{V}$ (mag) & 2.0\tablenotemark{1} \\
$R_\mathrm{in}$ (AU) & 65 \\
$R_\mathrm{out}$ (AU) & 170 \\
$i$ (degree) & 40 \\
$p$ & 1.5 \\
$q$ & 0.5 \\
\enddata
\label{tab:sedfit1}
\tablenotetext{1}{\citet{bib:hughes1994}}
\end{deluxetable}

\begin{deluxetable}{cc}
\tabletypesize{\small}
\tablecaption{Best-fit parameters in the SED fitting}
\tablewidth{0pt}
\tablehead{
\colhead{Parameter} &
\colhead{Value}
}
\startdata
\multicolumn{2}{c}{Stellar parameters}\\
\hline
$R_\ast$ ($R_\sun$) & 1.29$\pm$0.03 \\
$T_\ast$ (K) & 4148$\pm$55 \\
$L_\ast$ ($L_\sun$) & 0.49$\pm$0.02 \\
\hline
\multicolumn{2}{c}{Cold outer disk} \\
\hline
$T_\mathrm{in}$ (K) & 32.5$\pm$3.9 \\
$\Sigma_\mathrm{in}$\tablenotemark{1} (g cm$^{-2}$) & 0.67$\pm$0.03 \\
$\beta$ & 0.5$\pm$0.1 \\
$M_\mathrm{disk}$\tablenotemark{1} (10$^{-3}$ $M_\sun$) & 2.4$\pm$0.8 \\
\enddata
\label{tab:sedfit2}
\tablenotetext{1}{The surface density and the disk mass are shown in the gas+dust density and mass by assuming the gas-to-dust mass ratio of 100.}
\end{deluxetable}

\subsection{Origin of the Hot Component in the Inner Hole of the Disk}\label{sec:sedimp}
The SED model analysis indicates that the presence of a hot component inside the dust disk with a temperature of $\sim$180 K, which is responsible for the SED peak at $\sim$20 $\micron$.
There are two possible origins for such a hot component: a localized self-luminous emitting body (i.e., circumplanetary disk) or an inner warm structure of the disk \citep{bib:wolf2005}.
In this section, we estimate the size and mass of the hot component for both cases.\par

In the case of the circumplanetary disk, we can estimate the radius from $\Omega_\mathrm{c}$ because it is usually optically thick \citep{bib:dangelo2003}.
Assuming the circumplanetary disk is parallel to the parent circumstellar disk, the solid angle of $1\times10^{-16}$ str corresponds to the radius of 0.3 AU, or 64 $R_\sun$.
On the other hand, the Hill radius of a putative planet around \object{Sz 91} is expressed by
\begin{equation}
 R_\mathrm{Hill} = 1.1 \times \frac{a}{10\ \mathrm{AU}} \biggl( \frac{m}{1\ M_\mathrm{J}}\biggr)^{1/3} \ \mathrm{[AU],}
\end{equation}
where $m$ and $a$ are the mass and orbital radius of the planet, respectively.
The radius of the circumplanetary disk inferred from the SED fitting is significantly smaller than $R_\mathrm{Hill}$ at 3--65 AU from the star, which is consistent with the theoretical expectation for a circumplanetary disk \citep[e.g.,][]{bib:tanigawa2012}.
However, our ADI observation with the Subaru telescope could not examine the presence of such a companion planet due to a low rotation angle of $\sim16\degr$.\par

For the second possibility, because the hot component could be fitted with a single temperature, it must be confined to a narrow width in radius.
If we extrapolate the temperature distribution of the outer disk determined in the SED fitting, the temperature of the hot component in the optically thick case (172 K) corresponds to a radius of 2.3 AU.
The best-fit $\Omega_\mathrm{c}=1\times10^{-16}$ str corresponds to a ring width of 0.01 AU, which is significantly narrow with respect to the ring radius.
In contrast, the optically thin condition is unlikely because the solid angle is $>4\times10^{-13}$ str which corresponds to a ring width of $>$53 AU.
Therefore we conclude that the other implication of the hot component is the optically thick ring at 2.3 AU, whose total mass is at least $6\times10^{-7}$ $M_\sun$.
Such an example of the optically thick ring around the transition disk has also been reported in \mbox{RX J1633.9-2442} \citep{bib:cieza2012a}.
The inner structure of \object{Sz 91} may be similar to that of \mbox{RX J1633.9-2442}, whereas \object{Sz 91} exhibits a larger inner hole and a lower disk mass.\par

\subsection{Velocity Structure of the Gas Disk: Model Calculation and Comparison with CO(3--2) Profile}\label{sec:linecalc}
The CO(3--2) image shown in Figure \ref{fig:sma_co_int} suggests the presence of a rotating gas disk around the star.
To reveal the disk rotation in detail, we calculated model spectra of CO(3--2) with a simple power-law disk model according to that reported by \citet{bib:kitamura1993}, and we compared the results with the observed CO(3--2) profile.

The following parameters of the model disk were estimated from the observed images and the SED fitting: the inner and outer disk radii of 65 and 420 AU, respectively; the temperature distribution of $T(r)=32.5\times(r/R_\mathrm{in})^{-0.5}$ K, and the surface density distribution of $\Sigma(r)=0.67\times(r/R_\mathrm{in})^{-1.5}$ g cm$^{-2}$.
The hydrostatic equilibrium is assumed along the vertical direction and the density distribution, $\rho(r,z)$, is therefore expressed by
\begin{equation}
\rho(r,z) = \rho(r,0) \exp \biggl[ - \frac{1}{2} \biggl( \frac{z}{H(r)} \biggr)^2 \biggr] \mathrm{,}
\end{equation}
where $H(r)$ is the scale height given by
\begin{equation}
H(r) = \sqrt{\frac{r^3 k_\mathrm{B}T(r)}{GM_\ast \mu m_\mathrm{H}}} \mathrm{.}
\end{equation}
Here, $G$ is the gravitational constant, $M_\ast$ is the stellar mass, $\mu$ is mean molecular weight, and $m_\mathrm{H}$ is the mass of H atom.
The density at the midplane is expressed by
\begin{equation}
\rho(r,0)=\frac{\Sigma(r)}{\sqrt{2\pi}H(r)} \mathrm{.}
\end{equation}
The gas motion is assumed to be Kepler rotation whose velocity field is written by
\begin{equation}
V(r) = \biggl( \frac{G M_\ast}{r} \biggr)^{0.5} \mathrm{.}
\end{equation}
The fractional abundance of CO with respect to H$_2$ is assumed to be $9\times10^{-5}$, which corresponds to a typical interstellar value \citep[e.g.,][]{bib:schloerb1984,bib:irvine1985}.
Because the density of the disk is at least $10^6$ cm$^{-3}$ at the midplane which is significantly higher than the critical density of CO(3--2) ($\sim10^4$ cm$^{-3}$), we assume local thermodynamic equilibrium (LTE).
Notably, the hot component in the dust hole was not included because its contribution was negligible ($\lesssim1\times10^{-4}$ Jy).\par

\begin{figure}
	\epsscale{1.0}
        \plotone{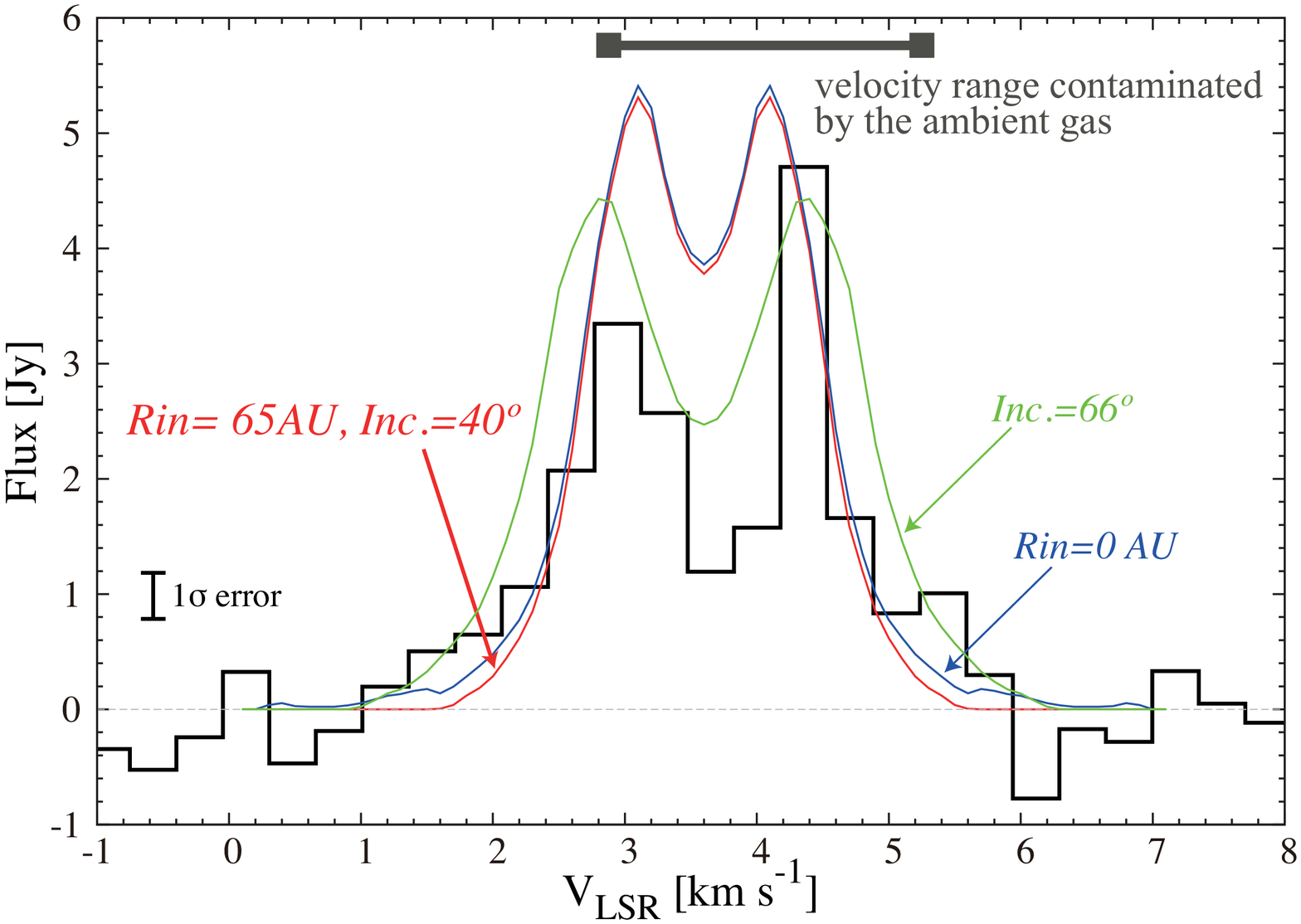}
	\caption{CO(3--2) spectrum integrated over a $5\arcsec \times 5\arcsec$ area centered at the stellar position (black line). The red line shows the calculated profile by a disk model with an inner radius of 65 AU and an inclination angle of 40$\degr$ from face-on (see text for the detailed description of the model). The blue line indicates the case of the inner radius of 0 AU. The green line indicates the model profile of the disk with a 65 AU inner radius and an inclination of 66$\degr$. The CO emission in the velocity range from 4 to 7 km s$^{-1}$ in $V_\mathrm{LSR}$ is probably contaminated by the ambient cloud emission, as shown by the gray line at the top of the panel.}\label{fig:sma_co_spec}
\end{figure}

Figure \ref{fig:sma_co_spec} shows the calculated CO spectra superimposed on the observed CO spectrum integrated over a $5\arcsec\times5\arcsec$ box centered at the stellar position.
We noted that the peak intensity of the calculated CO spectrum differed from the observed value.
However, the discrepancy is not significant because the disk emission around the stellar LSR velocity is probably contaminated by the ambient cloud emission.
Therefore, we focused on the emission at the blue-shifted side ($V_\mathrm{LSR}<2.9$ km s$^{-1}$).
The model profile in the $i=40\degr$ case agrees with the observed data, confirming the validity of the disk parameters derived from the dust disk.
Notably, the inclination angle of 66$\degr$ determined by the 345 GHz image did not fit well the CO line shape, indicating that the inclination of the dust disk estimated from the NIR scattered light is more plausible.

The presence of the gas inside the inner edge of the dust disk could not be confirmed from our data set.
Although we calculated the model spectrum for $R_\mathrm{in}=$0 AU as shown in Figure \ref{fig:sma_co_spec}, the difference is within the $1\sigma$ uncertainty.
The presence of the gas disk in the inner hole is supported by the fact that \object{Sz 91} shows the mass accretion onto the star, and thus, higher spatial resolution imaging is required to reveal the inner structure of the gas disk.\par

\section{Summary}\label{sec:summary}
We present the results of the aperture synthesis 345 GHz continuum and CO(3--2) line emission observations with SMA toward a transition disk object in Lupus, \object{Sz 91}.
Furthermore, a high resolution image of the polarized intensity at the $K_s$ band obtained with the Subaru Telescope is also presented.
The transition disk around \object{Sz 91} has been directly resolved and imaged in this study.
The disk parameters are derived and the structure and evolutional phase of the \object{Sz 91} transition disk are discussed.
The main results of our observations are summarized in the following points:\par

\begin{itemize}
 \item{Our high-resolution imaging revealed a dust disk around \object{Sz 91} with inner and outer radii of 65 and 170 AU, respectively, and an inclination angle of 40$\degr$. Furthermore, the Kepler rotating gas disk with a radius of 420 AU was imaged in the CO(3--2) line.}
 \item{Model analysis of the SED of \object{Sz 91} was performed by using a simple power-law disk model. We determined that the observed SED can be reproduced well by the combination of a cold disk and a hot component in the inner hole of the disk. The total H$_2$ mass of the cold disk is estimated to be $2.4\times10^{-3}$ $M_\sun$ if the canonical gas-to-dust mass ratio of 100 is adopted. The disk mass is significantly higher than those of other class III sources in nearby star forming regions; however,the disk mass is one of the lowest masses among the currently known transition disks.}
 \item{We determined that the hot component can be expressed by a single temperature gray body of $\sim180$ K. Although the hot component could not be resolved by our observations, its origin is either a localized self-luminous emitting body (i.e., a Jovian mass protoplanet with a circumplanetary disk) or an optically thick ring in the inner hole of the disk at 2.3 AU.}
 \item{Our results confirm the previous results such that the disk structure of \object{Sz 91} is consistent with that of an ongoing giant planet forming disk. In particular, the relatively large inner hole and lower disk mass indicate that the transition disk of \object{Sz 91} is probably in a stage of nearly completing planet formation. \object{Sz 91} will be a crucial target for investigating the evolution of transition disks and the planetary formation process. In the near future, our proposed study with the Atacama Large Millimeter/submillimeter Array (ALMA) will provide a new insight into the planet formation process.}
\end{itemize}

\acknowledgements
We are grateful for the ASTE and AzTEC staff for the operation and maintenance of the observation instruments.
A part of this work was conducted as the Observatory Project of ``SEEDS: Strategic Explorations of Exoplanets and Disks with Subaru'' supported by the MEXT Grant-in-Aid for Scientific Research on Priority Areas.
This work is partially supported by JSPS KAKENHI Grant Numbers 24103504 (T.T.) and 23103004 (M.M.).
J.C. gratefully acknowledges support from NSF grant AST-1009203.
The Submillimeter Array is a joint project between the Smithsonian Astrophysical Observatory and the Academia Sinica Institute of Astronomy and Astrophysics and is funded by the Smithsonian Institution and the Academia Sinica.

\end{document}